\documentclass[5p,twocolumn,sort&compress,times,preprint,normalem]{elsarticle}

\usepackage{graphics}
\usepackage{graphicx}

\usepackage{amsfonts, amsthm, amssymb}
\usepackage{color}
\usepackage{ascmac}
\usepackage{url}
\usepackage{ulem}
\usepackage{bm}
\usepackage{bbold}
\usepackage{amsmath}
\usepackage{algorithm}
\usepackage{algpseudocode}
\usepackage{braket}
\usepackage{multirow}
\usepackage{subfiles}
\usepackage{cprotect}
\usepackage{epsfig}
\usepackage{xspace}
\usepackage{tipa}
\usepackage{newverbs}

\newverbcommand{\verbred}{\color{black}\bgroup}{\egroup}

\newcommand{\thoshi}[1]{#1}

\newcommand{\tyoshimi}[1]{#1}

\newcommand{\rev}[1]{#1}

\newcounter{bla}

\journal{Computer Physics Communications}

\begin{document}

\begin{frontmatter}



\title{sim-trhepd-rheed -- Open-source simulator of  total-reflection high-energy positron diffraction (TRHEPD) and reflection high-energy electron diffraction (RHEED)}


\author[a]{Takashi Hanada}
\author[b]{Yuichi Motoyama}
\author[b]{Kazuyoshi Yoshimi}
\author[c,d,e]{Takeo Hoshi\corref{author}}
\cortext[author] {Corresponding author.\\\textit{E-mail address: hoshi@tottori-u.ac.jp} }

\address[a]{Institute for Materials Research, Tohoku University, 2-1-1 Katahira, Aoba-ku, Sendai 980-8577, Japan}
\address[b]{Institute for Solid State Physics, University of Tokyo, 5-1-5 Kashiwa-shi, Chiba 277-8581 Japan}
\address[c]{Department of Mechanical and Physical Engineering, Tottori University,  4-101 Koyama Minami, Tottori-shi, Tottori 680-8552, Japan}
\address[d]{Advanced Mechanical and Electronic System Research Center, Faculty of
Engineering, Tottori University, 4-101 Koyama Minami, Tottori-shi, Tottori 680-8552, Japan}
\address[e]{Slow Positron Facility, Institute of Materials Structure Science, High Energy Accelerator Research Organization (KEK), Oho 1-1, Tsukuba, Ibaraki, 305-0801, Japan}

\begin{abstract}
The present paper reports sim-trhepd-rheed (STR), an open-source simulator of total-reflection high-energy positron diffraction (TRHEPD) and reflection high-energy electron diffraction (RHEED) experiments \tyoshimi{which are} used for atom-scale surface structure determination of a material. 
\tyoshimi{Diffraction data obtained by these experiments are analyzed by comparison with numerical simulations.}
The \tyoshimi{STR} simulator is used for the analysis of experimental diffraction data by simulating the rocking curve from a given trial surface structure \tyoshimi{by solving} the partial differential equation of the dynamical quantum diffraction theory for positron or electron wavefunctions.
\thoshi{Using the obtained surface structure, electronic structure, and} \tyoshimi{other physical quantities can be evaluated through first-principles calculations. For this purpose,
a} utility software was also developed in order to realize a \tyoshimi{first principles} calculation with the Quantum ESPRESSO suite.
\end{abstract}


\end{frontmatter}



{\bf PROGRAM SUMMARY}

\begin{small}
\noindent
sim-trhepd-rheed -- Open-source simulator of  total-reflection high-energy positron diffraction (TRHEPD) and reflection high-energy electron diffraction (RHEED) \\
{\em Authors:} Takashi Hanada, Yuichi Motoyama, Kazuyoshi Yoshimi, Takeo Hoshi.
\\
{\em Program title:} sim-trhepd-rheed \\
{\em Journal reference:}   \\
{\em Catalogue identifier:}                                   \\
{\em Program summary URL:} \\
https://github.com/sim-trhepd-rheed/sim-trhepd-rheed/\\
{\em Licensing provisions:} GNU General Public License v3.0\\
{\em Programming language:} Fortran~90, Python~3                                   \\
{\em Computer:} Any architecture\\
{\em Operating system:} Unix, Linux, macOS  \\
{\em RAM:} Depends on the number of variables \\
{\em Number of processors used:} Arbitrary 
\\
{\em External routines/libraries:} BLAS library,
LAPACK library\\
{\em Nature of problem:}
Partial differential equation in dynamical quantum diffraction theory \\ 
{\em Solution method:} Numerical solution by the multi-slice method for the partial differential equation\\
\rev{{\em Code Ocean capsule:} (to be added by Technical Editor)}
\\
\\

\end{small}

\section{Introduction \label{Sec:Intro}}

Reflection high-energy electron diffraction (RHEED) and 
total-reflection high-energy positron diffraction (TRHEPD) 
\cite{HUGENSCHMIDT_2016_SurfSciRep_rev, Fukaya_2018_JPHYSD, 
Fukaya_2019_Book_SurfaceStructure, Mochizuki_2016_PCCP_TiO2, ENDO_20208_Carbon} 
are experimental probes for surface structure
and form the foundation of surface science. 
Reflection high-energy electron diffraction has been used as a standard experimental probe for decades, and
TRHEPD is a novel experimental probe \cite{HUGENSCHMIDT_2016_SurfSciRep_rev, Fukaya_2018_JPHYSD, 
Fukaya_2019_Book_SurfaceStructure, Mochizuki_2016_PCCP_TiO2, ENDO_20208_Carbon}.
\tyoshimi{The experimental technique of} TRHEPD was first proposed in 1992 by Ichimiya \cite{ichimiya1992_TRHEPD} and 
was realized in a study in 1998 by Kawasuso and Okada \cite{kawasuso1998_prl}.
Following a period of initial development by the Kawasuso group, this technique has been actively developed in the last decade at \tyoshimi{large-scale experimental facilities} \thoshi{at} the Slow Positron Facility (SPF), Institute of Materials Structure Science (IMSS), High Energy Accelerator Research Organization (KEK).

The present paper reports that
we developed and recently released 
the \lq sim-trhepd-rheed' (STR) software package, which is 
a GPL-based open-source simulator of RHEED and TRHEPD.
The \tyoshimi{STR} simulator was originally developed by Takashi Hanada, one of the present authors. 
The calculation method used in \tyoshimi{STR} is 
the multi-slice method ~\cite{Ichimiya_1983_JJAP_RHEED_solver, ICHIMIYA_1987_SurfSciLett_OneBeam, Fukaya_2018_JPHYSD}.
The code was used for the data analysis of RHEED  
~\cite{RHEED_SURF_SCI_1993, RHEED_HIKITA_JVST_1993, RHEED-Kudo-SurfIntf-1994, RHEED-HANADA-SurfSci-1994, HANADA_PRB_1995,RHEED-YAMADA-PRL-1995,RHEED-OHTAKE-PRB-1999, RHEED-Miotto-ApplPhysLett-1999, RHEED-Ohtake-JCrysGrow-1999, RHEED-Ohtake-PRB60-1999, RHEED-RealTime-Ohtake-PRB60-1999, RHEED-Miotto-2000,RHEED-Ohtake-PRB-2001,RHEED-Ohtake-PRB-2002}
and was later extended for the data analysis of TRHEPD 
\cite{TANAKA2020_ACTA_PHYS_POLO, TANAKA_2020_Preprint, Hoshi_2021_SiON_Preprint}. 
Note that a simulation code with the multi-slice method 
was developed by Ichimiya \cite{Ichimiya_1983_JJAP_RHEED_solver, ICHIMIYA_1987_SurfSciLett_OneBeam}
and was later modified by \rev{his} colleagues. 
The \tyoshimi{STR} simulator was developed independently from scratch \rev{as a more versatile and flexible tool, which can treat any type of crystal structure and surface orientation. 
Type of surface space group can be selected to reduce the number of independent atomic coordinates in the unit mesh.
This is an indispensable feature to analyze efficiently surface structure.}
In addition, several useful \tyoshimi{scripts} 
are included in the STR package, such as a \tyoshimi{script} that enables \tyoshimi{first principles} calculation by Quantum ESPRESSO (\url{https://www.quantum-espresso.org/})
\cite{QUANTUM_ESPRESSO_2009}


The remainder of the present paper is organized as follows. First, an overview of the calculation method is given in Section \ref{Sec:Theory}. Next, the basic information of the software, such as installation and usage, is introduced in Section~\ref{Sec:Software}. Then, all parameters in input files are explained in detail with an example and the usage of two supporting tools are illustrated with another example in Section~\ref{Sec:Sample}. 
Two examples of comparison between the calculated and measured rocking curves are shown in Section~\ref{Sec:Comparison}.
Finally, we summarize the present paper in Section~\ref{sec:summary}. In addition, usage of the utility to prepare input files for Quantum ESPRESSO is introduced in Appendix A.


\begin{figure}[h]
\begin{center}
  \includegraphics[width=0.40\textwidth]{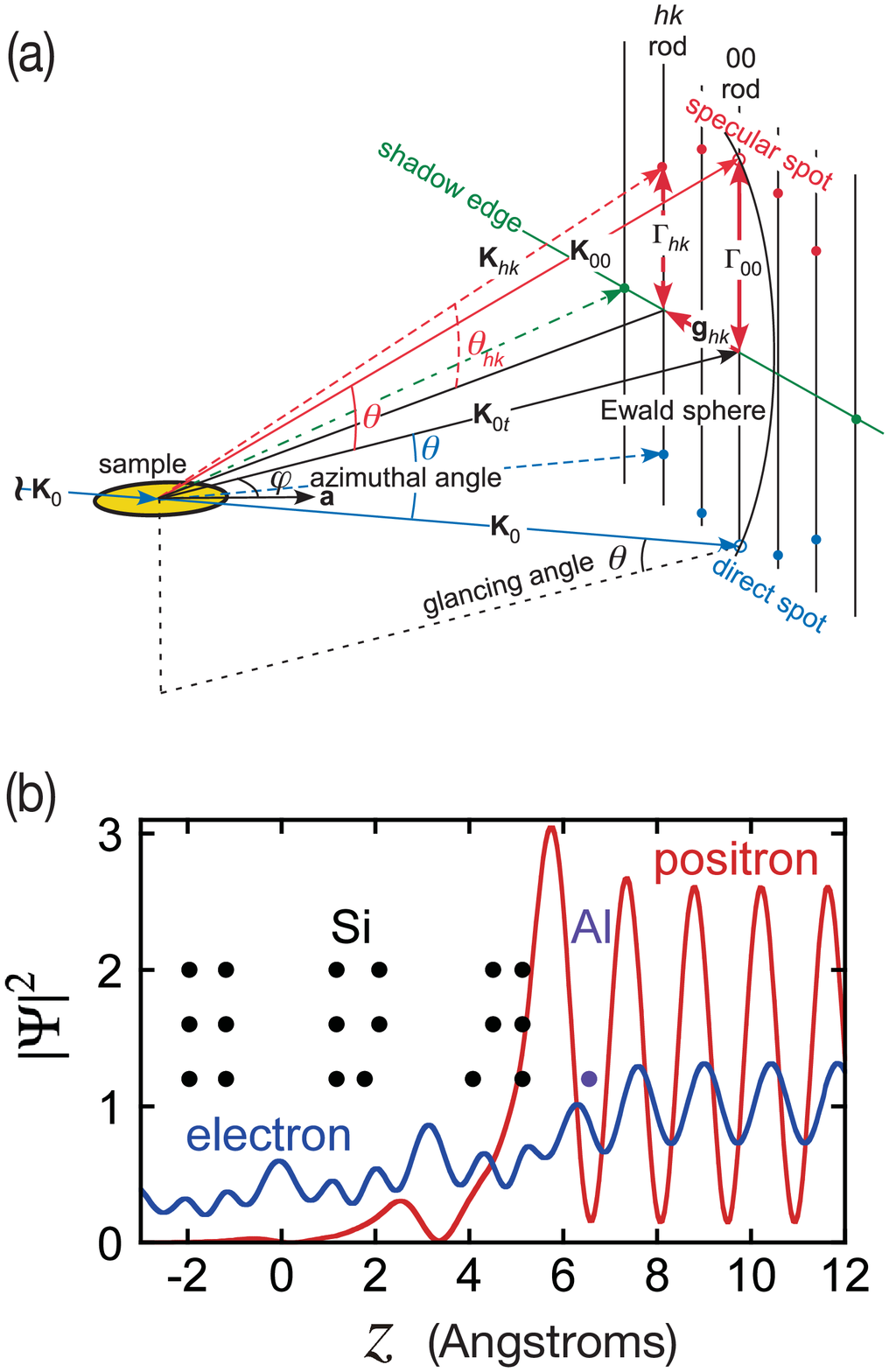}
\end{center}
\caption{(a) Schematic diagrams of \tyoshimi{the} TRHEPD and RHEED experiment. 
From conservation of the kinetic energy in vacuum, 
$|\bm{K}_{0}|$ = $|\bm{K}_{00}|$ = $|\bm{K}_{hk}| \equiv K$. 
Diffraction spots appear in the direction of the intersection point of the 
reciprocal-lattice rods and the Ewald sphere of radius $K$.
Here, $\bm{K}_{0t}$ varies with $\theta$, and the angle between 
$\bm{K}_{0t}$ and $\bm{g}_{hk}$ varies with $\varphi$.
(b) An example of   
\tyoshimi{the positron/electron} density along the surface-normal direction in the TRHEPD/RHEED experiment on Si(111)-($\sqrt{3} \times \sqrt{3}$)R30$^\circ$ surface with a T4-site Al adatom. In (b), the beam energy is 15 keV, the glancing angle $\theta$ is 2$^\circ$, and the one-beam condition is assumed. The atomic position and number of atoms in the unit mesh is indicated by the closed circles. The incident beam of unity amplitude and reflected beam form a standing wave just above the surface, where they overlap. The positron penetrates only a few layers below the surface at small $\theta$ due to the repulsive force from nuclei.
}
\label{FIG-TRHEPD-OVERVIEW}       
\end{figure}

\section{Overview of the proposed method \label{Sec:Theory}}

Figure \ref{FIG-TRHEPD-OVERVIEW}(a) illustrates
the common experimental setup of TRHEPD and RHEED. 
Positron and electron beams are used for TRHEPD and RHEED, respectively. 
From the experimental intensity distribution in a diffraction pattern, 
we obtain the rocking curve $I^{\rm (exp)}_{hk}(\theta)$, 
the glancing angle dependence of the intensity,
where the indices $(h, k)$ specify the surface reciprocal lattice vector and a particular spot on the screen. 
\rev{The} multi-slice method is based on 
the dynamical quantum diffraction problem to solve the wavefunction $\Psi(\bm{r})$
of the beams, the kinetic-energy and corresponding wave number $K$ in a vacuum of which are  given, in a two-dimensional periodic potential of the crystal surface. 
An example of the density of the particles under the one-beam condition, which is mentioned later, is shown in Figure \ref{FIG-TRHEPD-OVERVIEW}(b),
where the $z$ axis is chosen to be perpendicular to the surface.
The calculation method numerically solves
the partial differential equation (PDE) at a given glancing angle and azimuthal angle of the incident beam
\cite{Ichimiya_1983_JJAP_RHEED_solver} 
\begin{eqnarray}
\left( \Delta + K^2 -q  U(\bm{r})  \right) \Psi(\bm{r}) =   0, 
\label{EQ-PDE}
\end{eqnarray}
so as to obtain the intensity $I^{\rm (cal)}_{hk}(\theta)$.
The parameter $q$ indicates the sign of the particle \tyoshimi{($+/-$ for positron/electron).}
The crystal potential  $q U(\bm{r})$ \tyoshimi{, which is a complex function owing to the inelastic effect,} is determined by the lattice parameters, the atomic positions $\bm{r}_n$ of $n$-th atom in a unit mesh, and the atomic scattering factors corrected by mean-square displacements.
The database of the atomic scattering factors \cite{RHEED-Doyle-1968, RHEED-Peng-Micron-1999} 
is implemented in STR among the atomic numbers of 
$Z$=1 (H), 2 (He), ..., 98 (Cf). 

The two primitive translation vectors of the bulk truncated $1 \times 1$ surface
are denoted as $\bm{a}$ and $\bm{b}$,
and those of its reciprocal space are denoted as $\bm{a}^\ast$ and $\bm{b}^\ast$,
respectively. Then, two primitive translation vectors of 
a reconstructed surface structure can be written as 
\begin{eqnarray}
& & \bm{a}_s = n_{aa} \bm{a} + n_{ab} \bm{b} \label{EQ-VEC-AS} \\
& & \bm{b}_s = n_{ba} \bm{a} + n_{bb} \bm{b}. \label{EQ-VEC-BS}
\end{eqnarray}
The two primitive translation vectors
of its reciprocal lattice are \tyoshimi{given as }
$\bm{a}_s^\ast = (n_{bb} \bm{a}^\ast - n_{ba} \bm{b}^\ast)/s$
and 
$\bm{b}_s^\ast = (n_{aa} \bm{b}^\ast - n_{ab} \bm{a}^\ast)/s$, 
\tyoshimi{where $s=n_{aa} n_{bb} - n_{ab} n_{ba}$}. 
The area of the surface unit mesh is $\Omega = s ab \sin \gamma$, \tyoshimi{where $a$ and $b$ are the norms of $\bm{a}$ and $\bm{b}$, respectively, and $\gamma$ is the angle between these two vectors.} 
In addition, 
$\bm{a}^\ast = n_{aa} \bm{a}_s^\ast + n_{ba} \bm{b}_s^\ast$
and
$\bm{b}^\ast = n_{ab} \bm{a}_s^\ast + n_{bb} \bm{b}_s^\ast$
are easily confirmed.

The crystal potential $U(\bm{r})$ is periodic on the surface of the $x-y$ plane and can be
written in terms of the two-dimensional Fourier series
\begin{eqnarray}
U(\bm{r}_t, z) =   \sum_{h,k} U_{hk}(z) \exp \left( i \bm{g}_{hk} \bm{r}_t \right),  
\label{EQ-POT}
\end{eqnarray}
where $\bm{g}_{hk} = h \bm{a}^\ast + k \bm{b}^\ast$ is the surface reciprocal lattice vector,
and \thoshi{${\bm r}_t$} is the position vector in the $x-y$ plane.
The Fourier component $U_{hk}(z)$ is given as 
\begin{eqnarray}
U_{hk}(z) &=& \frac{4\pi}{\Omega}  \sum_{n} \left( 1 - iq \varepsilon_{m(n)} \right)
\beta_{n} \exp \left[ - 2 \pi i \left( h x_n + k y_n \right) \right] \nonumber \\
& \times & \sum_{j=1}^4 a_{jm(n)} \exp \left( - \frac{b_{jm(n)}+B_{m(n)}^{xy}}{16 \pi^2} |\bm{g}_{hk}|^2 \right) \nonumber \\
& \times & \sqrt{\frac{4\pi}{b_{jm(n)}+B_{m(n)}^{z}}} \nonumber \\ 
& \times & 
\exp \left[ - \frac{4\pi^2}{b_{jm(n)}+B_{m(n)}^{z}} \left( z - z_n\right)^2 \right], 
\label{EQ-U-FOURIER}
\end{eqnarray}
where $n$, $m(n)$, and $j$ are the index of atoms in the unit mesh, 
the index of atomic species of the atom $n$, and the index of the four component Gaussians, respectively. The  $a_{jm(n)}$ and $b_{jm(n)}$ are the shape parameters of the Gaussians composing the atomic scattering factor specified by $m(n)$ \cite{RHEED-Doyle-1968, RHEED-Peng-Micron-1999}.
In addition, $\beta_{n}$ is the site occupation ratio of atom $n$.  
The position of atom $n$ is $(\bm{r}_{tn}, z_n)$, where $\bm{r}_{tn} = x_n \bm{a} + y_n \bm{b}$. The absorption effect of the beams \tyoshimi{is taken into account through} 
the imaginary potential, which is simply assumed as $\varepsilon_{m(n)}$ times the real potential in terms of the individual contribution from atom $n$.
Lastly, $B_{m(n)}^{xy} = 8 \pi^2 \langle (u_{m(n)}^{xy})^2 \rangle$
and  
$B_{m(n)}^{z} = 8 \pi^2 \langle (u_{m(n)}^{z})^2 \rangle$
are, respectively, the in-plane and surface-normal B-factors, which are $8 \pi^2$ times the mean square displacements of the atomic species $m(n)$. 
In the bulk region, $\Omega$ is $ab \sin \gamma$, and $h$ and $k$ are integers.

\begin{figure}[h]
\begin{center}
  \includegraphics[width=0.40\textwidth]{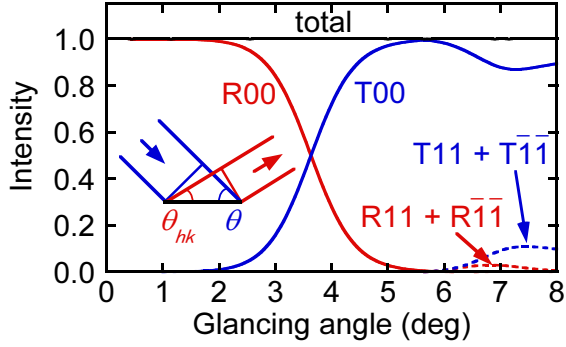}
\end{center}
\caption{The TRHEPD rocking curves of the reflected beams labeled \lq R00' and \lq R$11$+R$\bar{1}\bar{1}$', the transmitted beams labeled \lq T00' and \lq T$11$+T$\bar{1}\bar{1}$', and their total labeled \lq total' for monolayer graphene.
The intensity of the incident beam is redistributed without any loss in calculation if inelastic scatterings are ignored. 
}
\label{FIG-TRHEPD-GRAPHENE}       
\end{figure}

The wavefunction $\Psi(\bm{r})$ is written in the form of Bloch function
\begin{eqnarray}
\Psi(\bm{r}_t, z) = \sum_{h,k} c_{hk}(z) \exp \left[ i \left( \bm{K}_{0t} + \bm{g}_{hk} \right) \bm{r}_t \right],
\label{EQ-WAVE}
\end{eqnarray}
where $\bm{K}_{0t}$ is in-plane component of the incident wave vector shown in Fig. \ref{FIG-TRHEPD-OVERVIEW}(a).
In the vacuum above the surface and below the bottom of the sample, $U(\bm{r})$ is zero.
Then, $c_{hk}(z)$ is represented by incident and reflected plane waves above the surface
\begin{eqnarray}
c_{hk}(z) = \delta_{(hk)(00)} \exp \left( -i \Gamma_{hk} z \right) 
                      + R_{hk} \exp \left( i \Gamma_{hk} z \right)
\label{EQ-REFWAVE}
\end{eqnarray}
and the transmitted plane wave below the bottom
\begin{eqnarray}
c_{hk}(z) = T_{hk} \exp \left( -i \Gamma_{hk} z \right), 
\label{EQ-TRANSWAVE}
\end{eqnarray}
respectively, where $\delta_{(hk)(00)}$ is the Kronecker's delta 
and $\Gamma_{hk}^2 = K^2 - \left( \bm{K}_{0t} + \bm{g}_{hk} \right)^2$. 
If $\Gamma_{hk}^2 \geq 0$, $\Gamma_{hk} = K \sin \theta_{hk}$, as shown in Fig. \ref{FIG-TRHEPD-OVERVIEW}(a).
In addition, $R_{hk}$ (and $T_{hk}$ if the sample is thin enough) can be numerically calculated by the multi-slice method ~\cite{Ichimiya_1983_JJAP_RHEED_solver}.
Note that the direction of $+z$ in the present paper is opposite to that in ~\cite{Ichimiya_1983_JJAP_RHEED_solver}.
Finally, the reflection intensity of the $hk$ beam is evaluated as 
\begin{eqnarray}
I_{hk}^{\rm (cal,0)}(\theta) = |R_{hk}|^2 \frac{{\rm Re}(\Gamma_{hk})}{\Gamma_{00}} I_{0}(\theta),
\label{EQ-I-CALC}
\end{eqnarray}
where
\begin{eqnarray}
I_{0}(\theta) = 
\begin{cases}
\sin \theta / \sin \theta_c & (0 \leq \theta  \leq \theta_c)\\
1 & (\theta  > \theta_c),
\end{cases}
\label{EQ-I0}
\end{eqnarray}
$\sin \theta_c = \phi / w$, $\phi$ is the diameter of incident beam at the sample position, 
and $w$ is the width of the sample along $\bm{K}_{0t}$.
If the sample is thin enough, the intensity of the transmitted $hk$ beam is similarly evaluated as 
\begin{eqnarray}
I_{hk}^{\rm (trans)}(\theta) = |T_{hk}|^2 \frac{{\rm Re}(\Gamma_{hk})}{\Gamma_{00}} I_{0}(\theta).
\label{EQ-I-CALCT}
\end{eqnarray}
The factor of $I_{0}(\theta)$ is necessary owing to the grazing incidence condition of TRHEPD/RHEED and $I_{0}(\theta)$ approximates the fraction of the incident beam that irradiates the sample surface over the total incident beam. 
In the present code, however, $I_{0}(\theta) = \sin \theta$ is adopted to prevent the dependence on $\phi$ and $w$.
Instead, experimental intensity must be multiplied by $\sin \theta / \sin \theta_c$ above 
$\theta_c$ prior to data fitting. 
However, this correction is not necessary as long as the direct spot is observed.
For example, if $\phi$ is 1 mm and $w$ is 5 mm,
then $\theta_c$ is 11.5$^\circ$, which is above the scan range.
Moreover, a constant factor such as $\sin \theta_c$ can be ignored because scale factors of experimental and calculated intensities are adjusted during the fitting process.
Next, the factor of ${\rm Re}(\Gamma_{hk}) / \Gamma_{00}$ represents the ratio of the cross-sectional area between the reflected (transmitted) beam and the incident beam, as shown in the inset of Fig. \ref{FIG-TRHEPD-GRAPHENE}.
The flow rate of particles in a beam at constant $K$ is proportional to density multiplied by the area of the beam \cite{Ino-TRAXS-1996}.
More directly, ${\rm Re}(\Gamma_{hk}) / \Gamma_{00}$ is the ratio of the surface-normal component of momentum \cite{MAKSYM1981}.
The latter factor is particularly important when the $hk$ beam emerges from the shadow edge, which is shown in Fig. \ref{FIG-TRHEPD-OVERVIEW}(a), and ${\rm Re}(\Gamma_{hk})$ is small.
The product of the two factors is simplified as $\sin(\theta_{hk})$ if $\Gamma_{hk}^2 \geq 0$ and 0 if $\Gamma_{hk}^2 \leq 0$.

Figure \ref{FIG-TRHEPD-GRAPHENE} shows a confirmation that the total intensity labeled \lq total' of the reflected beams labeled \lq R00' and \lq R$11$+R$\bar{1}\bar{1}$' and the transmitted beams labeled \lq T00' and \lq T$11$+T$\bar{1}\bar{1}$' is equal to the intensity of the incident beam, which is assumed to be unity, i.e. $I_{0}(\theta) = 1$, for clarity. In this calculation, the TRHEPD intensities of the $00$, $11$, and $\bar{1}\bar{1}$ beams are calculated for a monolayer graphene without any inelastic scattering. The beam energy is 10 KeV, and the incident azimuth is $[1\bar{1}00]$. There is a wide total reflection region for the positron.

Under the one-beam condition, in which the incident azimuth is intentionally shifted from any low index direction \cite{ICHIMIYA_1987_SurfSciLett_OneBeam, Fukaya_2018_JPHYSD}, 
only the $(h,k)=(0,0)$ Fourier component is assumed to be non-zero\tyoshimi{. Thus, Eq.~(\ref{EQ-POT}) can be rewritten as}
\begin{eqnarray}
U(x,y,z) = U_{00}(z). 
\label{EQ-POT2}
\end{eqnarray}
The wavefunction can also be written as 
$\Psi({\bm r}) \equiv c_{00}(z) \exp \left( i \bm{K}_{0t} \bm{r}_t \right)$ 
and Eq. ~(\ref{EQ-PDE}) is reduced to a one-dimensional scattering problem 
\begin{eqnarray}
\left( \frac{d^2}{d z^2} + K^2 \sin^2 \theta -q U_{00}(z)  \right) c_{00}(z) = 0.
\label{EQ-ODE}
\end{eqnarray}
Under this condition, we can concentrate on only the surface-normal atomic coordinates as a first step to a full surface structure analysis.

The solution of the PDE gives the theoretical rocking curve $I^{\rm (cal,0)}_{hk}(\theta)$ as a function of the glancing angle $\theta$.
The rocking curve is broadened by the Gaussian function $g(\theta)$ with the standard deviation $\sigma$ 
by the convolution 
\begin{eqnarray}
& & I_{hk}^{\rm (cal)}(\theta) =  \int_{\theta_{\rm min}}^{\theta_{\rm max}}  
 I_{hk}^{\rm (cal, 0)}(\theta') g(\theta - \theta') d \theta', \label{EQ-CONV-1} \\
& & g(\theta) \equiv \frac{1}{\sqrt{2\pi}\sigma} \exp \left( -\frac{\theta^2}{2\sigma^2} \right). \label{EQ-CONV-2} 
\end{eqnarray}
In \tyoshimi{STR, $\sigma$ is determined through the input parameter of full width at half maximum \rev{(FWHM)} $\omega$, which satisfies $\exp(-\omega^2/8\sigma^2)=1/2$.}
\thoshi{
The parameter $\omega$ governs the broadening effect on the glancing angle $\theta$ due to the angular divergence of the incident beam and 
typically, $\omega$ is set as $0.5^\circ$ in the case of TRHEPD experiment.}
%

\section{Software \label{Sec:Software}}

This section details the usage of STR. 
The STR simulator consists of the three essential applications,  
\verb|bulk.exe|, \verb|surf.exe|, and \verb|make_convolution.py|, 
and the three optional utilities \verb|xyz.exe|, \verb|potcalc.exe|, and \verb|to_dft.py|. 
Except for the Python3 scripts, 
\verb|make_convolution.py| and \verb|to_dft.py|, the source code is written in Fortran90.
The source code is available on the github repository.
The numerical libraries of LAPACK and BLAS are also required.
The code was tested on Linux workstations and the
Fugaku (\url{https://www.r-ccs.riken.jp/en/fugaku/project}) and 
Oakforest-PACS (\url{http://jcahpc.jp/eng/ofp_intro.html}) supercomputers, among others. 
In the input file, 
we can specify the surface structure with the two-dimensional space groups. 
The index \verb|NSG| is shown in Table \ref{TABLE-SPACE-GROUP}.


\begin{table}[tb]
  \begin{center}
  \cprotect\caption{Index \verb|NSG| for the two-dimensional space groups. The origin of the primitive translation vectors is a center of the highest symmetry, if it exists. Otherwise, $\bm{a}$ of bulk and $\bm{a}_s$ of surface are on a reflection axis if NSG is 3 and 5, or on a glide axis if NSG is 4.}
  \begin{tabular}{|c|c|c|}  \hline
    index \verb|NSG| & symmetry \\  \hline
    1 & p1\\ \hline
    2 & p2\\ \hline
    3 & p1m1\\ \hline
    4 & p1g1\\ \hline
    5 & c1m1\\ \hline
    6 & p2mm\\ \hline
    7 & p2mg\\ \hline
    8 & p2gg\\ \hline
    9 & c2mm\\ \hline
    10 & p4\\ \hline
    11 & p4mm\\ \hline
    12 & p4gm\\ \hline
    13 & p3\\ \hline
    14 & p3m1\\ \hline
    15 & p31m\\ \hline
    16 & p6\\ \hline
    17 & p6mm\\ \hline
  \end{tabular}
  \label{TABLE-SPACE-GROUP}
  \end{center}
\end{table}

The default source code of STR 
is written for the TRHEPD experiment.
When using STR for the RHEED experiment,
the line
\lq \verb|ep='P'|'
should be modified to \lq \verb|ep='E'|'
in \verb|bulkm.f90|, \verb|surfm.f90| and \verb|U0.f90|. 
Hereafter, we show the calculated results for the TRHEPD case. However, input files are common to both TRHEPD and RHEED, although the optimum value of $\varepsilon_{m(n)}$ in Eq.~(\ref{EQ-U-FOURIER}) may be different for positrons and electrons.

\subsection{Installation \label{Sec:Inst}}

When one downloads the source code and 
executes the \verb|make| command on the \verb|src/| directory, 
the application binaries, 
\verb|bulk.exe|, \verb|surf.exe|, \verb|xyz.exe|, and \verb|potcalc.exe|,  
are built on the same directory. 
The Python3 scripts \verb|tool/make_convolution.py|, \verb|tool/todft/to_dft.py| also are available.
The built binaries and the Python3 script should be copied into \tyoshimi{the directory for which the path was set in the PATH environment variable.}

\subsection{Usage \label{Sec:Usage}}

\tyoshimi{The calculation using STR is performed by the following three processes.}
\tyoshimi{First,} the application \verb|bulk.exe| is executed 
so as to solve the PDE in the bulk region.
The required input file is \verb|bulk.txt|, which sets the atom positions in the bulk region. The output file is \verb|bulkP.b|, which is required in the next stage. 
\tyoshimi{Next,} the application \verb|surf.exe| is executed 
so as to solve the PDE in the surface region.
The required input file is \verb|surf.txt|,
which sets the atom positions in the surface region. 
The output file is \verb|surf-bulkP.s|, which contains
the rocking curve data $I_{hk}^{\rm (cal,0)}(\theta_j)$
for the preset angles $\{ \theta_j \}$.
\tyoshimi{Finally,} the Python3 script \verb|make_convolution.py| is executed so as to obtain the smoothed rocking curve data $I_{hk}^{\rm (cal)}(\theta_j)$ as the output file \verb|convolution.txt|.
\tyoshimi{For further calculations,} the software package contains optional utilities. The utilities \verb|xyz.exe| and \verb|potcalc.exe|
will be explained in Section~\ref{SEC:RESULT-GE001}
and the utility \verb|to_dft.py| in \ref{SEC:APP-TO-DFT}. 


\begin{figure}[h]
\begin{center}
  \includegraphics[width=0.40\textwidth]{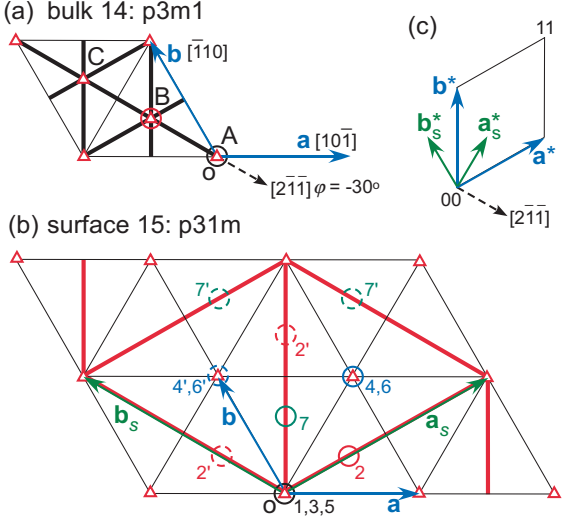}
\end{center}
\caption{Unit meshes and primitive translation vectors (solid arrows) of (a) the Si(111)-$1 \times 1$ bulk bilayer and (b) the Si(111)-($\sqrt{3} \times \sqrt{3}$)R30$^\circ$-Al surface. (c) Their reciprocal primitive translation vectors (solid arrows). Three-fold axes (triangles) and mirror planes (thick lines), which are normal to the surface, are shown, whereas glide planes are not. Open circles indicate atomic positions.
}
\label{FIG-ALT4-UNITMESH}       
\end{figure}

\section{\rev{Numerical} examples \label{Sec:Sample}}

The present package contains two examples in the \verb|sample/| directory. \tyoshimi{In this section, these examples are demonstrated with detailed explanations.}

\subsection{Si(111)-($\sqrt{3} \times \sqrt{3}$)R30$^\circ$ surface with a T4-site Al adatom}

The first example is the input data for 
Si(111)-($\sqrt{3} \times \sqrt{3}$)R30$^\circ$ surface with a T4-site Al adatom.
The incident azimuth is the $[2\bar{1}\bar{1}]$ symmetric direction shown in Fig. \ref{FIG-ALT4-UNITMESH}. Therefore, calculations with a sufficient number of reciprocal rods are necessary owing to the unavoidable multiple scatterings between the beams, the indices of which are $h = k$ ~\cite{RHEED-HANADA-SurfSci-1994, HANADA_PRB_1995}. 
The input data files, \verb|bulk.txt| and \verb|surf.txt|, 
are stored in the \verb|sample/T4Al_on_Si_111/| directory. 
The output files are stored 
in the \verb|sample/T4Al_on_Si_111/output/| directory. 

\subsubsection{Bulk part calculation}

\tyoshimi{The bulk part consists of the (111) bilayer of cubic silicon. In STR, the calculation of the bulk part is performed by specifying the following six types of parameters in the input file: (i) lattice parameters, (ii) indices $hk$ of Fourier components, (iii) setting of incident beam, (iv) atomic species, (v) atomic positions, and (vi) numerical condition.}
The input file \verb|bulk.txt| is as follows:
\verb|01: 3,3,1    ,NH,NK,NDOM,  ----- Si(111) bulk|\\
\verb|02: 13       ,NB|\\
\verb|03: 0        ,RDOM|\\
\verb|04: -6,-6,-5,-5,-4,-4,-3,-3,-2,-2,-1,-1,0,0,|\\
\verb|     1,1,2,2,3,3,4,4,5,5,6,6  ,(IH(I),IK(I))|\\
\verbred|05: 15,-30,-30,0,0.3,7,0.1|\\ 
\verb|     ,BE,AZI,AZF,DAZ,GI,GF,DG|\\
\verb|06: 0.1,100  ,DZ,ML|\\
\verb|07: 1      ,NELM|\\
\verbred|08: 14, 1.091, 0.1027    ,Si Z,da1,sap|\\
\verbred|09: 0.4782, 0.4782, 0.4782   ,BH,BK,BZ|\\
\verb|10: 14, 3.83966, 3.83966, 120, 3.13507, |\\
\verb|     -0.333333, 0.333333 |\\
\verb|     ,NSG,AA,BB,GAM,CC,DX,DY|\\
\verb|11: 2       ,NATM|\\
\verb|12: 1, 1, 0, 0, 1.17565 |\\
\verb|      ,IELM, ocr, X, Y, Z|\\
\verb|13: 1, 1, -0.333333, 0.333333, 1.95942 |\\
\verb|      ,IELM, ocr, X, Y, Z|\\
Hereafter, the line number of the text file
is indicated at the head of each line. 
\tyoshimi{For example, the content of the fourth line of} \verb|bulk.txt| \tyoshimi{is actually denoted as} \lq -6,-6,-5,-5,-4,-4,-3,-3,-2,-2,-1,-1,0,0, 1,1,2,2,3,3,4,4,5,5,6,6  ,(IH(I),IK(I))'.

\tyoshimi{The parameters for domains, symmetry, and lattice are specified in lines 1, 3, and 10 of the example file.}
A multi-domain case can be treated by specifying the number of the domains \verb|NDOM| (the third component of the first line) and 
\thoshi{the azimuthal-angle shift for each domain in the degree unit} \verb|RDOM| 
(the third line) .
\thoshi{In a multi-domain case, 
the azimuthal-angle shifts of the domains are written as separated by commas in the third line, for example,  \lq 0,120,240'.}
Hereafter, we assume the single-domain case in the present paper. Thus, \verb|NDOM| and \verb|RDOM| \tyoshimi{are set as $1$ and $0^{\circ}$, respectively.}
The \tyoshimi{details of treating} the multi-domain case are explained in the document file in the package. 
The two-dimensional space group (plane group) \tyoshimi{can be specified by} \verb|NSG| \tyoshimi{(the first component of the 10th line). The relation between} \verb|NSG| \tyoshimi{and the space group is shown in Table \ref{TABLE-SPACE-GROUP}. In this example, p3m1 is selected} as shown in Fig. \ref{FIG-ALT4-UNITMESH}(a). However, the result is the same even if p1 is selected because all atoms in the bulk are located on the three-fold axes.
The lengths of the two-dimensional primitive \tyoshimi{translation vectors, $\bm{a}$ and $\bm{b}$, and the angle between these two vectors $\gamma \equiv \cos^{-1} [(\bm{a} \cdot \bm{b})/(|\bm{a}||\bm{b}|)]$ can be specified by}  \verb|AA|, \verb|BB|, and  \verb|GAM| \tyoshimi{at the second, third, and fourth components, respectively, of the tenth line. 
In this example, $|\bm{a}| = |\bm{b}| = a_c / \sqrt{2}$ and $\gamma$ are chosen to be $3.83966$ \AA~ and $120^\circ$}, respectively, where $a_c$ is cubic lattice constant of bulk silicon. 
\tyoshimi{Note that} a convenient unit cell may not be the conventional unit cell for bulk because $\bm{a}$ and $\bm{b}$ have been set to be parallel to the surface. 
In order to form a compact unit cell, the primitive translation vectors of bulk are set as 
$\bm{a}_p = \bm{a}$, $\bm{b}_p = \bm{b}$, and
$\bm{c}_p = \delta x \bm{a} + \delta y \bm{b} + c_z \bm{e}_z$, 
where $c_z$ is the thickness of the unit cell and $\bm{e}_z$ is the unit vector in the $+z$ direction normal to the surface.
In this example, a diamond-type (111)-truncated structure consists of the three types of bilayers (\lq ABC'), and a unit cell of the present context consists of one bilayer of these three types of bilayers.
The in-plane shift of the stacked bilayers in upward direction $\delta x \bm{a} + \delta y \bm{b}$ can be chosen as $\overrightarrow{\mathrm{AB}} = \overrightarrow{\mathrm{BC}}$ in Fig. \ref{FIG-ALT4-UNITMESH}. 
Accordingly, in the tenth line, \verb|CC| = $c_z$ is set as $a_c / \sqrt{3} = 3.13507$ \AA \, at the fifth component, and \verb|DX| = $\delta x$ and \verb|DY| = $\delta y$ are set as $-1/3$ and $1/3$ at the sixth and seventh components, respectively.
The volume of the unit cell $a_c^3 /4$, which contains two atoms, is confirmed to have the same atomic density as the face-centered cubic cell.
As a side note, the primitive translation vectors of the reciprocal lattice are  
$\bm{c}_p^\ast = 2\pi/c_z \bm{e}_z$,
$\bm{a}_p^\ast = \bm{a}^\ast - \delta x \bm{c}_p^\ast$, and 
$\bm{b}_p^\ast = \bm{b}^\ast - \delta y \bm{c}_p^\ast$.
Although $\bm{c}_p$ is oblique, $\bm{c}_p^\ast$ is parallel to the reciprocal lattice rod, which is normal to the surface because both $\bm{a}_p$ and $\bm{b}_p$ are in the $x-y$ plane.
On the other hand, $\bm{a}_p^\ast$ and $\bm{b}_p^\ast$ are out of the $x-y$ plane.
However, this is not a problem because the first terms of these expressions determine the position of the reciprocal lattice rod to be $\bm{g}_{hk} = h \bm{a}^\ast + k \bm{b}^\ast$ in the $x-y$ plane
and the second terms of these expressions determine the starting point of the $|\bm{c}_p^\ast|$-interval bulk Bragg-peak positions along the rod ~\cite{RHEED-HANADA-SurfSci-1994, HANADA_PRB_1995}.
Furthermore, $|\bm{c}_p^\ast|$ is $2\pi\sqrt{3}/a_c$, which corresponds to the distance between bulk 000 and 111 reciprocal lattice points.

The parameters $(h, k)$ for the Fourier components in  Eq.~(\ref{EQ-POT}) \tyoshimi{are specified in lines 1, 2, and 4.}
The number of Fourier components \verb|NB| is chosen to be $13$ in the second line. 
A parameter set $(h,k)$ is specified 
by $(h,k)=(\verb|IH|/\verb|NH|,  \verb|IK|/\verb|NK|)$, 
where \verb|IH|, \verb|NH|, \verb|IK|, \verb|NK| are integer parameters. 
The parameters \verb|NH| and \verb|NK| are both chosen to be $3$ 
at the first and second components of the first line, 
and the \verb|NB|=13 sets of (\verb|IH|, \verb|IK|) are
chosen by the fourth line, as 
(\verb|IH|, \verb|IK|) =  (-6,-6),   (-5,-5), (-4,-4),  (-3,-3),   (-2,-2), (-1,-1), 
(0,0), (1,1), (2,2), (3,3), (4,4), (5,5), (6,6). Consequently, the Fourier component parameters are  
$(h_1,k_1)$=(-2,-2), $(h_2,k_2)$=(-5/3,-5/3),
$(h_3,k_3)$=(-4/3,-4/3),
$(h_4,k_4)$=(-1,-1),
$(h_5,k_5)$=(-2/3,-2/3),
$(h_6,k_6)$=(-1/3,-1/3),
$(h_7,k_7)$=(0,0),
$(h_8,k_8)$=(1/3,1/3),
$(h_9,k_9)$=(2/3,2/3),
$(h_{10},k_{10})$=(1,1),
$(h_{11},k_{11})$=(4/3,4/3),
$(h_{12},k_{12})$=(5/3,5/3),
$(h_{13},k_{13})$=(2,2). 
The locations of the (0,0) and (1,1) rods are indicated in Fig. \ref{FIG-ALT4-UNITMESH}(c).

The parameters for the incident beam are written in the fifth line for
the beam energy, \verb|BE|=15 eV, 
the initial azimuthal angle \verb|AZI|= -30$^{\circ}$ shown in Fig. \ref{FIG-ALT4-UNITMESH}(a),
the final azimuthal angle \verb|AZF|=-30$^{\circ}$,
the azimuthal angle step \verb|DAZ|= 0$^{\circ}$,
the initial glancing angle \verb|GI|= \rev{0.3}$^{\circ}$, 
the final glancing angle \verb|GF|= 7$^{\circ}$, and 
the glancing angle step \verb|DG|= 0.1$^{\circ}$. 

\tyoshimi{The parameters for the atomic species are specified in lines 7, 8 and 9.}
\tyoshimi{In line 7, the number of atomic species } \verb|NELM| \tyoshimi{is specified as 1 in the bulk section.}
In the present context, one \lq atomic species' indicates the parameter set of the scattering properties.
\tyoshimi{Therefore, when the same chemical species has different scattering properties, this species should be treated as a different atomic species $m(n)$.} 
Refer to Eq.~(\ref{EQ-U-FOURIER}) for the symbols shown here.
One atomic species is specified by the following six parameters \tyoshimi{listed in the eighth and ninth lines}: 
(i) the atomic number, \verb|iz|=14 for silicon, specified in the first component, which determines parameters $a_{jm(n)}$ and $b_{jm(n)}$,
(ii) \verb|da1|=\rev{1.091}, a correction term for parameter $a_{1m(n)}$ of Reference \cite{RHEED-Doyle-1968} (or $a_{4m(n)}$ of Reference \cite{RHEED-Peng-Micron-1999}), i.e., $a_{1m(n)}$ - \verb|da1| (or $a_{4m(n)}$ - \verb|da1|) is used in place of $a_{1m(n)}$ (or $a_{4m(n)}$),
(iii) \verb|sap| = $\varepsilon_{m(n)}$ = \rev{0.1027}, and 
(iv-vi) (\verb|BH, BK, BZ|) = ($B_{11m(n)}$, $B_{22m(n)}$, $B_{33m(n)}$) =\rev{(0.4782, 0.4782, 0.4782)},
where $B_{33m(n)}$ is $B_{m(n)}^{z}$ in Eq.~(\ref{EQ-U-FOURIER}).
The off-diagonal components of the B-factor are assumed to be 
$B_{12m(n)}$ = -$\left( B_{11m(n)} B_{22m(n)} \right)^{1/2} \cos \gamma$ 
and $B_{13m(n)}$ = $B_{23m(n)}$ = 0.
The in-plane B-factors must be isotropic, $B_{11m(n)}$ = $B_{22m(n)}$ = $B_{m(n)}^{xy}$, in most cases, as in Eq.~(\ref{EQ-U-FOURIER}). This is because symmetry operations are limited only to the in-plane atomic positions and those for anisotropic vibrations are not yet implemented in this code.
If NSG in Table \ref{TABLE-SPACE-GROUP} is less than 10, however, then the anisotropic in-plane B-factors are allowed under the assumption of the above-mentioned off-diagonal components because the anisotropy is preserved after the symmetry operations of these plane groups. 
Note that, in the calculated rocking curves, peak positions usually shift with \verb|da1| and peak widths usually increase with \verb|sap|. 
Here, \verb|sap| is inversely proportional to the square root of the beam energy \cite{Radi-1970}.

The parameters for \tyoshimi{the independent atoms in a bulk unit mesh are specified in lines 11, 12, and 13.}
The number of atoms \verb|NATM| in the unit is chosen to be $2$ in the 11th line. 
\tyoshimi{The parameters for the first atom are chosen in the 12th line for the atomic species} \verb|IELM|, which corresponds to $m(n)$ in Eq.~(\ref{EQ-U-FOURIER}), the occupation ratio \verb|ocr| = $\beta_{n}$, the atom position in the $x-y$ plane \verb|X| = $x_n$, \verb|Y| = $y_n$, and the atom position in the $z$ direction, \verb|Z| = $z_n$. 
In this code $\bm{a}$, $\bm{b}$, and $\bm{e}_z$ are the basis vectors in the bulk unit mesh 
and the atom position is specified by $x_n \bm{a} + y_n \bm{b} + z_n \bm{e}_z$. 
This is because the symmetry elements are normal to the surface, as shown in Fig. \ref{FIG-ALT4-UNITMESH}.
\verb|IELM| indicates the \verb|IELM|-th species among the list of the \verb|NELM| species. 
\tyoshimi{In this example, each parameter for the first atom 
shown by the circle at A  in Fig. \ref{FIG-ALT4-UNITMESH}(a) 
is set to} \verb|IELM|=1, \verb|ocr|=1.0, \verb|X|=0.0, \verb|Y|=0.0, and \verb|Z|= $\sqrt{3} a_c /8$ = 1.17565 \AA, respectively. 
The parameters for the second atom, shown by the circle at B in Fig. \ref{FIG-ALT4-UNITMESH}(a), are chosen in the 13th line in the same manner.
The second atom is located out of the primitive cell to align with the 'ABC' stacking sequence.
Obviously, translation by $\bm{a}$ and $\bm{b}$ causes no change in Eq.~(\ref{EQ-U-FOURIER}), because $h$ and $k$ are integers in bulk. 

The parameters for the numerical computation are written in the sixth line for \verb|DZ|=0.1 and \verb|ML|=100. 
The parameter \verb|DZ| is the thickness of the \lq slice', the interval of the discretized grid in the $z$ direction, and is written in angstroms. 
The parameter \verb|ML| is the cut-off number of the bulk unit layers. The number of layers is increased until the reflections from the bulk converge, but the repetition is terminated after ML layers.

The bulk part calculation is carried out by
\begin{verbatim}
$ bulk.exe
\end{verbatim}
and the output file \verb|bulkP.b|, a binary file, is generated. The output file is required in the surface part calculation, as explained in \tyoshimi{Section \ref{sec:4.1.2}.}

\subsubsection{Surface part calculation}\label{sec:4.1.2}

\tyoshimi{The calculation of the surface part is performed by specifying the four types of parameters in the input file: (i) atomic species, (ii) surface lattice parameters, (iii) atomic positions, and (iv) numerical condition. The input file of the surface part calculation} \verb|surf.txt| \tyoshimi{is as follows:}
\verb|01: 3      ,NELM,  -------- |\\
\verb|      Si(111)-root3 x root3-R30-Al T4|\\
\verbred|02: 13,-0.8458, 0.03746    ,Al Z,da1,sap|\\
\verbred|03: 0.7173, 0.7173, 0.7173     ,BH,BK,BZ|\\
\verbred|04: 14,1.091, 0.1027       ,Si Z,da1,sap|\\
\verbred|05: 0.7173, 0.7173, 0.7173     ,BH,BK,BZ|\\
\verbred|06: 14,1.091, 0.1027       ,Si Z,da1,sap|\\
\verbred|07: 0.4782, 0.4782, 0.4782     ,BH,BK,BZ|\\
\verb|08: 15,2,1,-1,1, 1.7, 0.333333,-0.333333 |\\ 
\verb|      ,NSGS,msa,msb,nsa,nsb,dthick,DXS,DYS|\\
\verb|09: 7      ,NATM|\\
\verbred|10: 1,1   0.000000   0.000000   6.587046|\\
\verbred|11: 2,1   0.643677   0.321838   5.150511|\\
\verbred|12: 3,1   0.000000   0.000000   4.042023|\\
\verbred|13: 3,1   1.000000   1.000000   4.503919|\\
\verbred|14: 3,1   0.000000   0.000000   1.801735|\\
\verbred|15: 3,1   1.000000   1.000000   2.084300|\\
\verbred|16: 3,1   0.333333   0.666667   1.142910|\\
\verb|17: 1      ,WDOM|\\

The parameters for the atomic species are written 
in the same manner as in the bulk part. 
The number of the atomic species in the surface part is chosen to be 
\verb|NELM|=3 in the \tyoshimi{first} line. 
For each atomic species, the six parameters,
\verb|iz|, \verb|da1|, 
\verb|sap|, \verb|BH|, \verb|BK|, and \verb|BZ|,
are written in the three pairs of lines from the \tyoshimi{second} line to the \tyoshimi{seventh} line.
\tyoshimi{In this example, t}he first atomic species is aluminum, and the second and third atomic species are silicon with different B-factors. 

The parameters for the surface symmetry, lattice, and domains \tyoshimi{are specified in lines 8 and 17.
In the eighth line, the parameters for the space group and the two-dimensional lattice vectors for the reconstructed surface, $\bm{a}_s \equiv n_{aa} \bm{a} + n_{ab} \bm{b}, \bm{b}_s\equiv n_{ba} \bm{a} + n_{bb} \bm{b}$ in Eqs.~(\ref{EQ-VEC-AS}) and (\ref{EQ-VEC-BS}), are specified. Note that the space group of the surface structure can be different from that of the bulk structure,
because of the surface reconstruction}, as shown in Fig. \ref{FIG-ALT4-UNITMESH}.
The space group is chosen to be p31m by the first component of the eighth line, \verb|NSGS|=15, 
according to Table \ref{TABLE-SPACE-GROUP}.
Some of the bulk symmetry elements are lost in the surface, i.e., the surface has a lower symmetry than bulk.
Here, \tyoshimi{$n_{aa}, n_{ab}, n_{ba},$ and $n_{bb}$ are specified by} \verb|msa|, \verb|msb|, \verb|nsa|, and \verb|nsb| in the second, third, fourth, and fifth components, respectively, of the eighth line. \tyoshimi{In this example, these parameters are chosen to be} \verb|msa|=2, \verb|msb|=1, \verb|nsa|=-1, \verb|nsb|=1.
The parameters \verb|DXS| and \verb|DYS| of the seventh and eighth components in the line specify the translation in which
the atomic coordinates of the surface parts are translated by 
\verb|DXS| $\bm{a}$ + \verb|DYS| $\bm{b}$. 
The translation is sometimes required so that the positions on the $x$-$y$ plane should be matched between the bulk part described by \verb|bulk.txt| and the surface part described by \verb|surf.txt|. 
In this example, \verb|DXS| and \verb|DYS| are chosen to be $1/3$ and $-1/3$, respectively. 
Parameter \verb|WDOM| is the weight of a domain and is chosen to be $1$ in the 17th line. This parameter should be specified for each domain in a multiple-domain case,
for example, '1,1,1'. 

The number of the independent atoms \verb|NATM| is chosen to be seven in the ninth line. 
The parameters for \tyoshimi{each atomic position} are written from the tenth to 16th lines. The parameters for the seven atoms are written in the same manner as those in the bulk part.
The zero point of the surface $z$-coordinate corresponds to the top of the bulk unit at $c_z$ = \verb|CC|.
The multiplicity of the first, third, and fifth atoms on the three-fold axis and three mirror planes is one as shown in Fig. \ref{FIG-ALT4-UNITMESH}(b).
The second and seventh atoms are multiplied by the three-fold rotation, and translated to create the two atoms shown by dashed circles labelled 2' and 7', respectively.
The fourth and sixth atoms are multiplied by the central mirror plane to create the atom shown by the dashed circle labelled 4' and 6'.
If whole surface layers shift (\verb|DXS|-1) $\bm{a}$ + \verb|DYS| $\bm{b}$, then the seventh atom is located just above the second atom in the bulk indicated by the circle at B in Fig. \ref{FIG-ALT4-UNITMESH}(a). 
The parameters for the numerical computation are \tyoshimi{specified at the sixth component in the eighth line.}
The parameter \verb|dthick| \tyoshimi{specifies} the thickness, in angstroms, above the topmost surface atom, where the potential has a tail toward the vacuum region. 
The thickness can be determined with the aid of \verb|potcalc.exe| explained in Section~\ref{SEC:RESULT-GE001}.
In this example, \verb|dthick| is chosen to be $1.7$.

The surface part calculation is carried out by 
\begin{verbatim}
$ surf.exe
\end{verbatim}
and the file \verb|surf-bulkP.s| \tyoshimi{is output}. 


The convolution calculation in Eq.~(\ref{EQ-CONV-1})
\tyoshimi{can be performed}
by the Python3 script \verb|tool/make_convolution.py|.
The input file is \verb|surf-bulkP.s|, and 
the full width at half maximum of the Gaussian function is chosen to be $\omega = 0.5^\circ$.
The procedure is carried out by 
\begin{verbatim}
$ python3 make_convolution.py \
   --filename surf-bulkP.s --omega 0.5 
\end{verbatim}
Here, the letter \lq \verb|\|' indicates that the real command line continues.
\tyoshimi{Then,} the file \verb|convolution.txt| \tyoshimi{is output. This file contains the rocking curve data}
$I_{hk}^{\rm (cal)}(\theta)$ for the glancing angle $\theta$ \tyoshimi{defined in the fifth line in} \verb|bulk.txt| and the thirteen index sets of $(h,k)$ \tyoshimi{defined in the fourth line in} \verb|bulk.txt|. 
At each line of \verb|convolution.txt|, 
the glancing angle $\theta $ and the thirteen rocking curve data $\{ I_{h_i k_i}^{\rm (cal)}(\theta) \}_{i=1,2,...,13}$ are listed in order. 

\begin{figure}[h]
\begin{center}
  \includegraphics[width=0.45\textwidth]{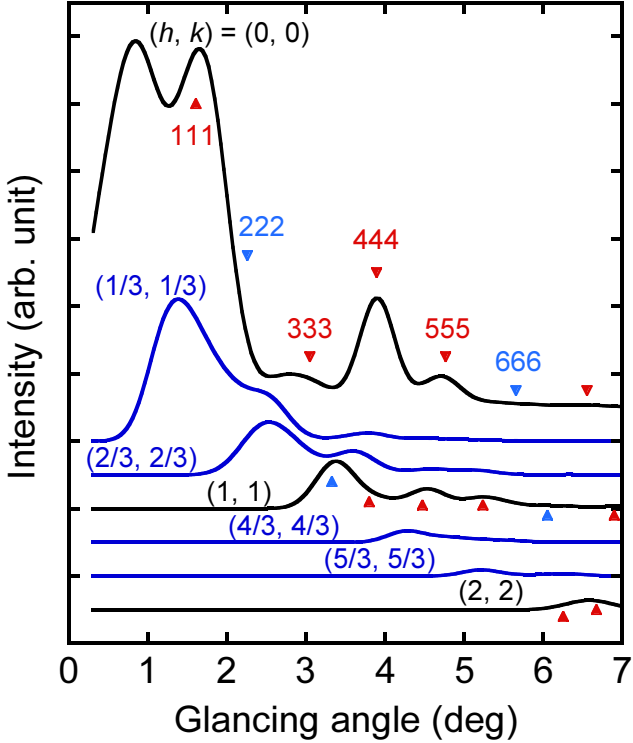}
\end{center}
\caption{Examples of TRHEPD rocking curves $I_{hk}(\theta)$
in a many-wave condition
for the Si(111)-\rev{($\sqrt{3} \times \sqrt{3}$)R30$^\circ$} surface with a T4-site Al adatom 
with $(h,k)=(0,0), (1/3,1/3), (2/3,2/3), (1,1), (4/3,4/3), (5/3,5/3), (2,2)$.
Estimated positions of Bragg reflections from bulk Si are indicated by triangles along integer-order rods for peak assignment.}
\label{FIG-THREPD-AlT4}       
\end{figure}

Figure \ref{FIG-THREPD-AlT4} shows the calculated 
TRHEPD rocking curves $I_{hk}^{\rm (cal)}(\theta)$ for 
$(h,k)=(0,0), (1/3,1/3), (2/3,2/3), (1,1), (4/3,4/3), (5/3,5/3), (2,2)$. 
Note that the 00-spot intensity, $I_{00}^{\rm (cal)}(\theta)$, is larger than the other spot intensities, as usual in the TRHEPD results.
The glancing angles corresponding to bulk Bragg reflections are estimated, and indicated by triangles.
Refraction of the beams at surface is taken into account by a model that $K^2$ in vacuum jumps to $K^2-qU_0$ in material \cite{Fukaya_2018_JPHYSD}, where the mean inner potential $U_0$ is assumed to be 8 eV.
The smaller effective $U_0$ for positron than that for electron, 12 eV for Si, agrees with a sense that the positrons are distributed at low potential regions owing to the repulsive force from nuclei as shown in Fig. \ref{FIG-TRHEPD-OVERVIEW}(b). 
In this example, the calculated peak positions in the integer-order rocking curves and the estimated angles of the Bragg reflections are fairly consistent at relatively large $\theta$ and $\theta_{hk}$.
The 222 and 666 reflections are forbidden in bulk.
However, notable modifications due to surface structure are seen at small $\theta$ or $\theta_{hk}$.
Moreover, the intensity of the peak assigned as 444 is much larger than those of 333 and 555. 
In bulk Si, squared norm of the crystal structure factor of 444 is only twice that of 333 and 555.
Therefore, surface structure analysis is possible by fitting the height and shape of the peaks with experimental rocking curves.
The fractional-order intensity curves originate from the surface structure and are crucial to surface structure analysis.

\subsection{Ge(001)-c(4 $\times$ 2) surface  \label{SEC:RESULT-GE001}}

The second example is the input data for 
Ge(001)-c(4 $\times$ 2) surface structure.
The measurement is set to be the one-beam condition.
The results are shown in our previous paper \cite{TANAKA_2020_Preprint}.
The present section details the usage of the two optional utilities, 
\verb|xyz.exe| and \verb|potcalc.exe|.

\begin{figure}[h]
\begin{center}
  \includegraphics[width=0.45\textwidth]{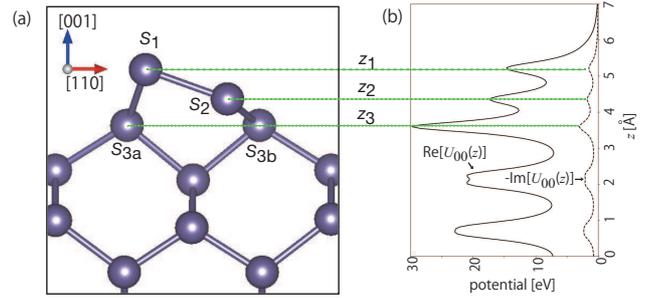}
\end{center}
\caption{
(a) Side view along the $[\bar{1} 1 0]$ direction
for a part of the Ge(001) surface,
in which only the asymmetric surface dimer (S$_1$, S$_2$) and several subsurface layers are shown. 
(b) Real and imaginary parts of the scattering potential 
averaged over the $x$-$y$ plane $({\rm Re}[U_{00}(z)], {\rm Im}[U_{00}(z)])$ 
converted into eV units are indicated by the solid and dashed lines, respectively, for the structure of (a). 
}
\label{FIG-SURFACE-GE001}       
\end{figure}

\begin{figure}[h]
\begin{center}
  \includegraphics[width=0.45\textwidth]{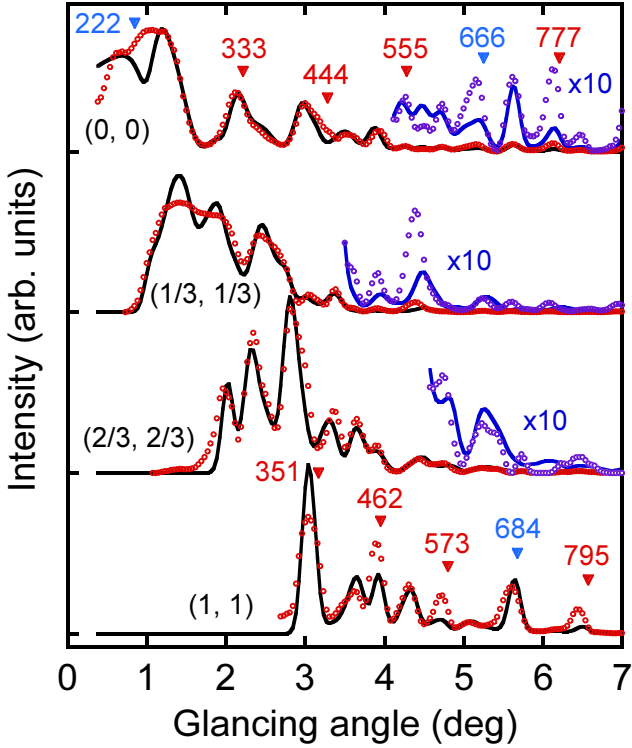}
\end{center}
\caption{\rev{Comparison between calculated (line) and experimental (circle) RHEED rocking curves for the Si(111)-($\sqrt{3} \times \sqrt{3}$)R30$^\circ$ surface with a T4-site Al adatom.} }
\label{FIG-RHEED-AlT4}       
\end{figure}

The utility \verb|xyz.exe| generates the atom position data into the xyz-format file,
and is used, for example, when visualizing the atomic structure is desired, 
as in Fig.~\ref{FIG-SURFACE-GE001} (a), 
and to confirm whether an intended structure is obtained from the input files of bulk and surface. 
The usage is explained as follows.
When one executes \verb|xyz.exe| in the directory 
that contains \verb|bulk.txt| and \verb|surf.txt|, 
the message
\begin{verbatim}
# of a-units, b-units & bulk layers ? :
\end{verbatim}
appears and requires the three integer parameters $(u_a, u_b, u_c)$, 
so as to determine the region 
of the generated structure file.
For example, we enter
\begin{verbatim}
1 1 2
\end{verbatim}
for the case of $(u_a, u_b, u_c)=(1,1,2)$.
The two-dimensional lattice vectors of the generated structure file
are $\bm{a}_{\rm file} \equiv u_a \bm{a}_{\rm s}$ and
 $\bm{b}_{\rm file} \equiv u_b \bm{b}_{\rm s}$.
The thickness of the bulk region in the generated structure file 
is set to be $u_c$ times the bulk unit. 
The generated structure file is written in the standard xyz format
but the last two lines of the file indicate
the additional information of the two-dimensional lattice vectors
$\bm{a}_{\rm file}$ and $\bm{b}_{\rm file}$.
An example of the generated structure file is
\verb|sample/Ge_001_c4x2/output/surf-bulk.txt|.
The file was generated with the parameters
of $(u_a,u_b,u_c)=(1,1,2)$.  
The last two lines of the file 
\begin{verbatim}
    16.00267     0.00000
     0.00000     8.00134
\end{verbatim}
indicate the two-dimensional periodic cell vectors,  
$\bm{a}_{\rm file}$=(16.00267 \AA, 0) 
and 
$\bm{b}_{\rm file}$=(0, 8.00134 \AA)

The utility \verb|potcalc.exe| outputs only the (0,0) component potential 
$V_{00}(z) = \hbar^2 U_{00}(z) / 2m$,
where $m$ is the mass of positron or electron, even if many-beam input data is used. 
The potential $V_{00}(z)$ is complex and the real and imaginary parts
are drawn in Fig.~\ref{FIG-SURFACE-GE001} (b).
The kinetic energy of the beam in a material is that in a vacuum minus the real part of the potential. 
In the case of RHEED, the sign of the real part is flipped. 
Therefore, the positron is decelerated and the electron is accelerated in a material.
The usage is explained as follows.
When \verb|potcalc.exe| is executed, 
the message
\begin{verbatim}
 # of bulk-unit repetition ? 
\end{verbatim}
appears and requires one integer parameter, 
so as to define the lower limit of $z$ in the generated numerical data of the potential $V_{00}(z)$. 
For example, we enter
\begin{verbatim}
 3
\end{verbatim}
and the output file will contain the numerical data of the potential $V_{00}(z)$
for the surface region and the bulk region,
where the bulk region consists of three bulk-unit repetitions.


\section{\rev{Comparison with experimental data} \label{Sec:Comparison}}

\rev{This section demonstrates 
the comparison between experimental and numerical rocking curves, where 
the numerical rocking curve is obtained from a trial data set of the atomic positions.
It is noted that, in general, the determination of surface structures may be carried out by the collaboration with other experimental and theoretical methods for suggesting a structure and/or confirming the suggested structure, as explained in the review paper \cite{Fukaya_2018_JPHYSD} and references therein. 
The structures in this section were obtained from fitting procedures with initial guesses.  }

\subsection{\rev{RHEED for Si(111)-($\sqrt{3} \times \sqrt{3}$)R30$^\circ$ surface with a T4-site Al adatom} \label{Sec:Comparison:rheed}}

\rev{
Figure. \ref{FIG-RHEED-AlT4} shows 
the calculated (line) and experimental (circle) RHEED rocking curves for the surface structure described in Section~\ref{Sec:Sample}. 
The mean experimental intensity in Ref. \cite{HANADA_PRB_1995} is normalized to be unity for each beam. The calculated curves are convoluted with a Gaussian function, FWHM of which is $0.16^\circ$. Then scale of each calculated curve is set to minimize the residual sum of squares (RSS) with observed one.
The glancing angles corresponding to bulk Bragg reflections are indicated by triangles, where the mean inner potential $U_0$ is assumed to be -12 eV. The Bragg reflections shift toward lower angles compared to Fig. \ref{FIG-THREPD-AlT4} owing to the opposite sign of $U_0$. 
The calculated curves are different from the ones in Ref. \cite{HANADA_PRB_1995} mainly because
$I_{hk}^{\rm (cal,0)}(\theta) = |R_{hk}\Gamma_{hk}/\Gamma_{00}|^2 I_{0}(\theta)$
was used in Ref. \cite{HANADA_PRB_1995} instead of Eq.~(\ref{EQ-I-CALC}).
$\Gamma_{00}$ and $\Gamma_{hk}$ are proportional to $\sin \theta$ and $\sin \theta_{hk}$, respectively. Therefore, peaks appear at nearly the same positions but intensity ratios between the peaks are modified particularly near the emergence angles of the non-specular beams. }
\rev{Another point to note is that the imaginary crystal potentials calculated in Ref. \cite{Radi-1970} were converted to Doyle-Turner-type imaginary atomic scattering factors in Ref. \cite{HANADA_PRB_1995}.
However, number of crystals calculated in Ref. \cite{Radi-1970} is limited.
Therefore, STR employs a much simpler parameter of} \verbred|sap| \rev{, which can fairly approximate the absorption effect of beams by adjusting its value.  
The atomic coordinates of the surface atoms and} \verbred|da1|\rev{,} \verbred|sap|\rev{,} \verbred|BH|\rev{,} \verbred|BK|\rev{, and} \verbred|BZ| \rev{of the bulk and surface atoms are newly optimized using STR with an aid of MINPACK subroutine LMDIF as in Ref. \cite{SAM-CTR-2018}. 
The optimal parameters are listed as the input data in Section~\ref{Sec:Sample}.
The parameter} \verbred|da1| \rev{of Al is optimized independently in the present work. 
This improves the agreement between the calculation and the experiment considerably.
The negative value of} \verbred|da1| \rev{suggests that Al adatom has a slight positive charge \cite{RHEED-Doyle-1968}.
It is assumed that the B-factors are isotropic and those of Al and first layer Si are 1.5 times as large as bulk one.
The estimated B-factor of bulk Si is consistent with results of an \textit{ab initio} calculation and X-ray diffraction \cite{Malica_B-factor}.
Finally $\theta_c$ is set at $7^\circ$ because RSS monotonously decreases with increasing $\theta_c$ and almost constant above $5^\circ$.
The rocking curves in Fig. \ref{FIG-RHEED-AlT4} have more peaks than those in Fig. \ref{FIG-THREPD-AlT4} because multiple scattering effect is stronger in RHEED than in TRHEPD. 
This is most probably because electrons penetrate into the material deeper than positrons as shown in Fig. \ref{FIG-TRHEPD-OVERVIEW}(b). At each glancing angle of bulk Bragg reflection of the 11 beam, the 00 beam have a resonant peak in RHEED. }


\subsection{\rev{TRHEPD for Si$_4$O$_5$N$_3$ / 6H-SiC(0001)-($\sqrt{3} \times \sqrt{3}$) R30$^\circ$ surface} \label{Sec:Comparison:trhepd}}

\rev{
The STR simulator was used in 
a recent data analysis of TRHEPD experiment \cite{Hoshi_2021_SiON_Preprint}.
In Figure 2 of Ref.\cite{Hoshi_2021_SiON_Preprint},
the numerical rocking curve data was compared 
with the experimental one 
for Si$_4$O$_5$N$_3$ / 6H-SiC(0001)-($\sqrt{3} \times \sqrt{3}$) R30$^\circ$ surface. 
The trial atomic positions were optimized by an automated optimization procedure 
\cite{TANAKA2020_ACTA_PHYS_POLO, TANAKA_2020_Preprint},  
till the numerical rocking curve reproduces well the experimental one. 
The software framework for the optimization algorithm was released later, 
by some of the present authors (Y. Motoyama, K Yoshimi, T. Hoshi), 
as an open-source software package}
\verbred|2DMAT| \rev{\cite{REF-2DMAT} ver. 0.1 in February 2021.
The automated optimization procedure is realized by installing both 2DMAT and STR. }

\section{Summary}
\label{sec:summary}

The sim-trhepd-rheed (STR) ver.1 open-source simulator was developed 
for data analysis of experimental results for total-reflection high-energy positron diffraction (TRHEPD) 
and reflection high-energy electron diffraction (RHEED).
The two diffraction experiments determine the atomic positions of the material surface. 
The foundation of the simulator is the numerical solution 
of the dynamical quantum diffraction theory for positron or electron wavefunctions.
The present paper contains 
a brief overview of the theory, 
the usage of the code, and several examples. 
The package also contains useful utilities, such as the utility for density functional theory (DFT) calculation.
The package is now available online under GNU General Public License v3.0.

\section*{Acknowledgements}

We would like to acknowledge support for the code development 
from the \lq Project for advancement of software usability in materials science (PASUMS)' 
by the Institute for Solid State Physics, University of Tokyo. 
The present research was supported in part by  
a Grant-in-Aid for Scientific Research (KAKENHI) from the Japan Society for the Promotion of Science 
(19H04125,20H00581). 
The present code was tested on 
the Fugaku supercomputer through the HPCI project (hp210083) and 
the Oakforest-PACS supercomputer as part of the  
Interdisciplinary Computational Science Program in the Center for Computational Sciences, University of Tsukuba.
 The present code was tested also 
on the supercomputers of the Supercomputer Center, Institute for Solid State Physics, University of Tokyo 
and at the Academic Center for Computing and Media Studies, Kyoto University.
We would like to thank Masaharu Hidaka and Hayato  Ichinose for fruitful discussions on the code.

\appendix

\section{Utility for density-functional theory calculation \label{SEC:APP-TO-DFT}}

The present appendix details 
the utility \verb|to_dft.py| available in the directory \verb|tool/todft| of the package.
The utility 
enables DFT calculation by Quantum ESPRESSO
\tyoshimi{by reading the output file in STR}.
The utility is a Python3 script that requires  
the atomic simulation environment (ASE) (\url{https://wiki.fysik.dtu.dk/ase/})
\cite{ASE_2017}.
The utility also requires the Python modules
\verb|numpy|, \verb|scipy|, \verb|matplotlib|, and \verb|toml|.

The usage of the utility is explained below
with an example of a Si(001)-$2 \times 1$ surface system. 
Two input files are available in the directory \verb|tool/todft/sample/Si001/|.
One input file is an xyz-formatted structure file, 
\verb|surf_Si001.xyz|.
The content of the file is 
\begin{verbatim}
01: 12
02: Si(001) surface test 
03: Si    1.219476    0.000000    9.264930
04: Si    6.459844    0.000000    9.987850
05: Si    1.800417    1.919830    8.404650
06: Si    5.878903    1.919830    8.404650
07: Si    3.839660    1.919830    7.155740
08: Si    0.000000    1.919830    6.900440
09: Si    3.839660    0.000000    5.743910
10: Si    0.000000    0.000000    5.597210
11: Si    1.919830    0.000000    4.321250
12: Si    5.759490    0.000000    4.321250
13: Si    1.919830    1.919830    2.963750
14: Si    5.759490    1.919830    2.963750
\end{verbatim}
The xyz-formatted file 
can be used but the information of the two-dimensional
lattice vectors, $\bm{a}_{\rm file}$and $\bm{b}_{\rm file}$, is ignored.

The other input file \tyoshimi{written in the toml format} \verb|input.toml| \tyoshimi{is prepared for setting the numerical parameters of DFT calculation. }
The content of the file is 
\begin{verbatim}
01: [Main]
02: input_xyz_file = "surf_Si001.xyz"
03: output_file_head = "surf_Si001_output"
04: [Main.param]
05: z_margin = 0.001
06: slab_margin = 10.0
07: [Main.lattice]
08: unit_vec = [[7.67932, 0.00000, 0.00000], 
    [0.00000, 3.83966, 0.00000]]
09: [H_term]
10: r_SiH = 1.48 #angstrom
11: theta = 109.5 #H-Si-H angle in degree
12: [ASE]
13: solver_name = "qe"
14: kpts = [3,3,1]        
    # sampling k points (Monkhorst-Pack grid)
15: command = "mpirun -np 4 ./pw.x -in 
    espresso.pwi > espresso.pwo"
16: [Solver]
17: [Solver.control]
18: calculation='relax' 
     # 'scf','relax','bands',...
19: pseudo_dir='./'     
     # Pseudopotential directory
20: [Solver.system]
21: ecutwfc = 20.0        
     # Cut-off energy in Ry
22: [Solver.pseudo]
23: Si = 'Si.pbe-n-kjpaw_psl.1.0.0.UPF'
24: H = 'H.pbe-kjpaw_psl.1.0.0.UPF'
\end{verbatim}
The second and third lines specify 
the filename of the input and output files. 
The fifth line specifies the margin value $ z_{\rm margin}$ in angstroms.
The value is used to extract the atoms in the bottom layer.
With the lowest $z$-coordinate $z_{\rm min} = {\rm min}_{i}z_i$, 
the atoms in 
$z_{\rm min} - z_{\rm margin} \le z \le z_{\rm min} + z_{\rm margin}$ 
will be extracted.
The sixth line specifies the slab margin $L_{\rm margin}$ in angstroms. 
If the $z$-coordinates of the atoms in the bottom and top layers are 
$z_{\rm min}$ and $z_{\rm max}$, respectively, then 
the slab thickness $L_{\rm slab}$, i.e., the periodic cell length on the $z$-axis, 
is given by $ L_{\rm slab} = z_{\rm max}-z_{\rm min} + L_{\rm margin}$. 
The eighth line specifies
the two-dimensional lattice vectors
of $\bm{a}_{\rm file}$=(7.67932 \AA, 0) 
and $\bm{b}_{\rm file}$=(0, 3.83966 \AA).

As an optional function, 
the utility supports the generation of hydrogen-terminated models.
The present code supports 
the hydrogen-terminated models
for the present (001)-type slab of the diamond structure.
The hydrogen-terminated models are generated in the following manner.
The bottom layer atoms are removed, and H atoms are placed at the corresponding positions in order to create a model with the distance to the next layer atoms adjusted to a tetrahedral structure (for example, the distance to a silane molecule in the case of Si). 
All lines after line 12 are written for Quantum ESPRESSO.
The details are explained in the document of the STR package. 

The utility is executed by the following command:
\begin{verbatim}
$ python3 to_dft.py input.toml
\end{verbatim}
After finishing calculations, the following files are generated:
\begin{verbatim}
surf_Si001_output.xyz
surf_Si001_output.cif 
espresso.pwi
espresso.pwo
\end{verbatim}
The files \verb|surf_Si001_output.xyz| and \verb|surf_Si001_output.cif|
are the (hydrogen-terminated) structure files 
in the xyz-format and the crystallographic information file (CIF) format, respectively. 
The files \verb|espresso.pwi| and \verb|espresso.pwo|
are the input and output files for Quantum ESPRESSO. 
The three files \verb|surf_Si001_output.xyz|,
\verb|surf_Si001_output.cif|, and \verb|espresso.pwi| 
indicate the same structure data and 
the files in the xyz and CIF format
may be useful for visualization.  
If Quantum ESPRESSO is not installed or does not works properly, then
the file \verb|espresso.pwo| contains only error messages. 
In such a case, 
Quantum ESPRESSO can be performed on a different machine with the file \verb|espresso.pwi|.








\end{document}